\documentclass{du-journals}

\pdfoutput=1

\journal{Iranian Journal of Astronomy and Astrophysics}

\title{A study on Ca~\texttt{II} 854.2~nm emission in a sunspot umbra using a thin cloud model}

\author[1]{H. Hamedivafa}
\address[1]{Physics Department, Faculty of Science, Imam Khomeini International University,
Qazvin 34149-16818, Islamic Republic of Iran; email:
vafa@sci.ikiu.ac.ir}
\author[2]{M. Sobotka}
\address[2]{Astronomical Institute, Academy of Sciences of the Czech Republic (v.v.i.),
Fri\v{c}ova 298, CZ–25165 Ond\v{r}ejov, Czech Republic; email:
michal.sobotka@asu.cas.cz}
\author[3]{L. Bellot Rubio}
\address[3]{Instituto de Astrof\'{\i}sica de Andaluc\'{\i}a (CSIC), Apdo. 3004, 18080 Granada, Spain; email:
lbellot@iac.es}
\author[3]{S. Esteban Pozuelo~$^{4, }$}
\address[4]{Institute for Solar Physics, Department of Astronomy, Stockholm University, AlbaNova University
 Center, 106 91 Stockholm, Sweden; email: sara.esteban@astro.su.se}

\begin{document}

\begin{abstract}
In the present work, we introduce and explain a method of solution
of the radiative transfer equation based on a thin cloud model. The
efficiency of this method to retrieve dynamical chromospheric
parameters from Stokes~\textit{I} profiles of
Ca~\texttt{II}~854.2~nm line showing spectral emission is
investigated. The analyzed data were recorded with the Crisp Imaging
Spectro-Polarimeter (CRISP) at Swedish 1-m Solar Telescope on La
Palma on 2012 May 5 between 8:11 - 9:00~UT. The target was a large
decaying sunspot (NOAA~11471) at heliocentric position W~$15^\circ$
S~$19^\circ$. This sunspot has a large umbra divided into two umbral
cores (UCs). One of these UCs shows steady spectral emission in both
Ca~\texttt{II}~854.2~nm and H$\alpha$ lines, where downflows
prevail. The other UC shows intermittent spectral emission only in
Ca~\texttt{II}~854.2~nm, when umbral flashes are propagating. The
statistics of the obtained Doppler velocities in both UCs is
discussed.
\end{abstract}


\section{Introduction}\label{intro}
The sunspot chromosphere comprises a number of inhomogeneously
magnetized plasma features relevant to flows, waves, shocks. These
dynamical features affect the spectral absorption and emission
characteristics of the chromosphere of sunspots.

One of these dynamical features, an umbral flash (UF), is a sudden
brightening observed in the chromospheric core of the Ca~\texttt{II}
lines. UFs tend to appear with a periodicity of roughly 3~min
\cite{beckers69,navarro00a,voort03}. it seems that they are the
chromospheric counterpart of the photospheric oscillation and are
connected to the phenomenon known as running umbral/penumbral waves
(for a review see \cite{khomenko15}).

By LTE and non-LTE inversions using spectro-polarimetric
Ca~\texttt{II}~854.2~nm observations, de la Cruz Rodr\'{\i}guez
et~al. \cite{dela13aa} confirmed that UFs have very fine structure
with hot and cool material intermixed at sub-arcsecond scales. They
have found a temperature excess of about 1000~K and upflows of
4~km~s$^{-1}$ around the temperature minimum region. The
corresponding 2D temperature maps have shown a hot canopy above the
umbra during UFs, which was not observed in quiet phases. This
canopy coincides with areas showing enhanced line-core emission.

The first polarimetric observations of UFs
\cite{navarro00a,navarro00b} already revealed the occurrence of
``anomalous'' Stokes~\textit{V} profiles during the UF events.
Socas-Navarro et~al. \cite{navarro00a,navarro00b} explained
anomalous profiles by a two-component scenario: an unresolved
mixture in the horizontal direction where a certain filling factor
is occupied by a ``quiet'' component (no reversal and zero or
slightly downflowing velocity), while the rest is occupied by the
shock-wave (line core emission reversal and strong upflows).

Besides inversions based on one-, two- or multi-component models,
``cloud model'' represents an alternative spectral inversion
technique describing the transfer of radiation through cloud-like
structures located above the ``solar surface'', transmitting the
radiation coming from below according to their optical thickness and
source function. Beckers \cite{beckers64} introduced a simple
inversion technique, known in the literature as ``Beckers' cloud
model'', for inferring the physical parameters of the cloud.
Different solutions of the radiative transfer equation based on the
cloud model were reviewed by Tziotziou \cite{tzio07}. The
cloud-model method is much simpler than the complete calculation of
a model atmosphere based on the non-LTE inversion of chromospheric
spectral lines and it seems to be suitable for our purpose. The
presence of the hot canopy with line-core emission detected in
\cite{dela13aa} during UFs, i.e., a kind of cloud-like structure,
indicates that the cloud-model approach is viable.

We intend to obtain Doppler velocities in the chromosphere above a
sunspot umbra where we observe signatures of UFs (spectral
emissions) in the Ca~\texttt{II}~854.2~nm line using a suitable
cloud model describing umbral chromosphere. The cloud model provides
a good representation of the observed profiles and a reliable
quantitative description of the spatial distribution of the physical
parameters responsible for the observed intensity inhomogeneities
formed in the body of the cloud structure \cite{alissa90,tsiro93}.

The paper is arranged as follows: the data sets are described in
Sect.~\ref{dataset}. A general description of the chromospheric
spectral emissions observed in our data sets is provided in
Sect.~\ref{gen}. A detailed description of the cloud model and a
method to solve the radiative transfer equation based on a new idea
are provided in Sect.~\ref{bcm}. In Sect.~\ref{res} we describe the
results and the reliability of the obtained model parameters.
Finally, we summarize our conclusions in Sect.~\ref{conclusion}.

\section{Data Set}\label{dataset}

The data analyzed in this study were recorded with the Crisp Imaging
Spectro-Polarimeter (CRISP; \cite{scharmer06}) at the Swedish 1-m
Solar Telescope (SST; \cite{scharmer03}) on La Palma. The dataset
includes the simultaneous sequences of full Stokes profiles of
Ca~\texttt{II} infrared line at 854.2~nm, Fe~\texttt{I} at 617.3~nm,
and the Stokes~\textit{I} in H$\alpha$ acquired on 2012 May 5
between 8:11 - 9:00~UT (49~min). The target was a large decaying
sunspot (NOAA~11471) at heliocentric position W~$15^\circ$
S~$19^\circ$ equivalent to the heliocentric angle of 23~deg
($\mu=0.92$).

Data reduction was performed using the CRISPRED pipeline described
in \cite{dela15aa}. The Ca~\texttt{II}~854.2~nm observations are
affected by a backscatter problem. The CRISP acquisition system has
three back-illuminated Sarnoff cameras. Their quantum efficiency
decreases in the infrared wavelengths and, above approximately
700~nm, each CCD camera becomes partially transparent. Consequently,
images acquired at Ca~\texttt{II}~854.2~nm show a circuit pattern
that cannot be removed using standard dark and flat field
corrections. To solve this problem, the flat field process of the
Ca~\texttt{II}~854.2~nm was performed considering the aspects
discussed in \cite{dela13aa}.

The dataset consists of a time series of profile scans of the
mentioned three spectral lines with a cadence of 56.5~s (52~scans).
The Ca~\texttt{II} line was observed at 17 line positions with a
sampling of 0.01~nm between $\Delta\lambda=\pm$0.08~nm along with a
continuum sample at $\Delta\lambda=+0.24$~nm. The profile scans of
Fe~\texttt{I} line are sampled at 31 wavelengths ranging between
$\Delta\lambda=-30.8~$pm and $\Delta\lambda=+53.2$~pm around the
line center at 2.8~pm spacing. The scans of Stokes~\textit{I}
profiles of H$\alpha$ line are sampled at 21 wavelengths ranging
between $\Delta\lambda=\pm0.1$~nm around the line center with
0.01~nm spacing. The formation height of the iron line is in the
photosphere while both Ca~\texttt{II}~854.2~nm and H$\alpha$ are
chromospheric spectral lines, the former being formed at higher
layers \cite{cauzzi08}. According to Cauzzi, et al. \cite{cauzzi08},
the wings of Ca~\texttt{II} infrared line are formed in the middle
photosphere while its core comes from the middle chromosphere. After
data reductions and the co-aligning the scans of different spectral
lines, the FOV is reduced to $830\times865$ pixels
($49^{\prime\prime}\times51^{\prime\prime}$).

\section{General Properties and Chromospheric Emission}\label{gen}
\begin{figure} %Fig.~1
 \centerline{\includegraphics[width=\textwidth]{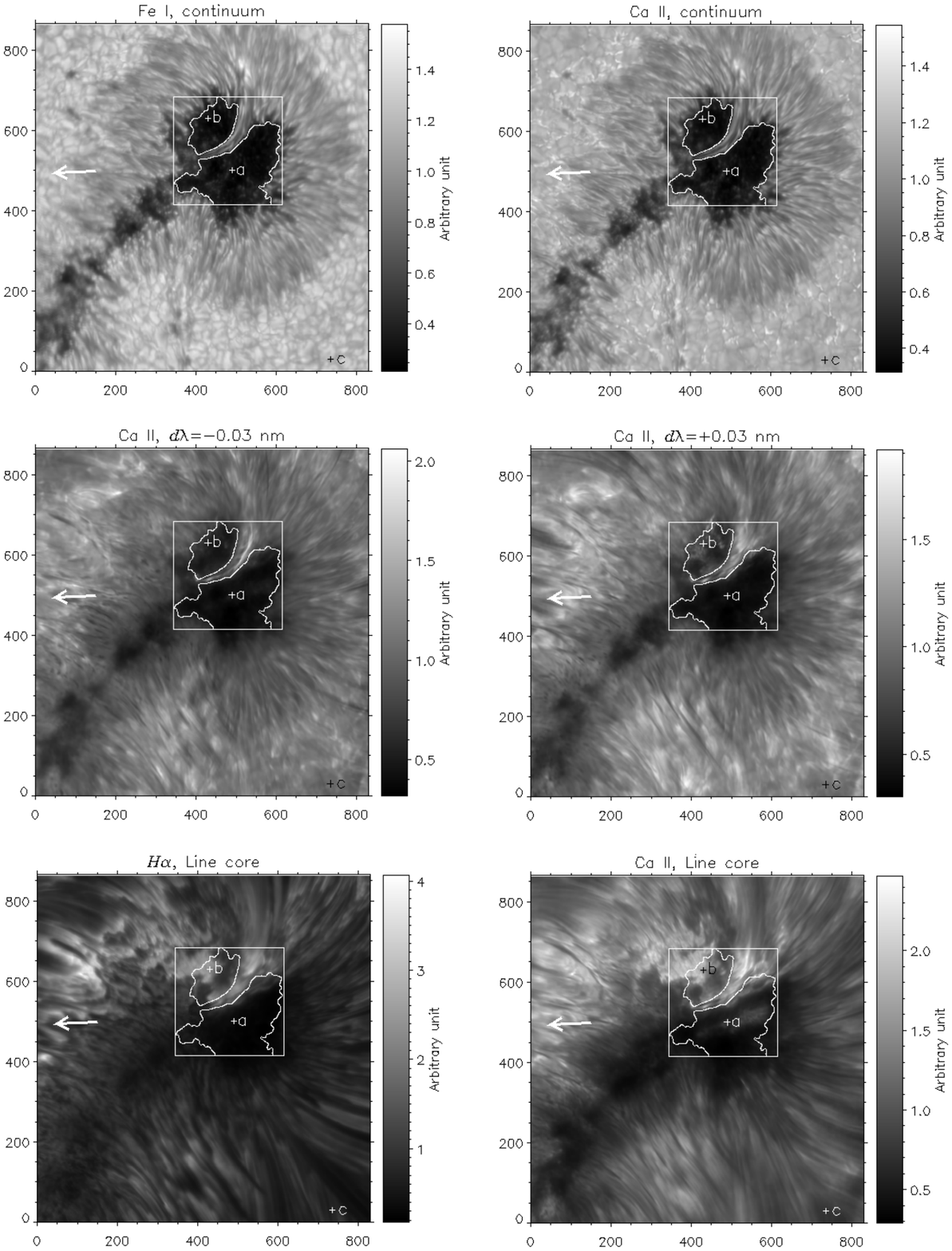}}
 \caption[]{Illustrations of filtergrams of three spectral lines (the fourth scan).
  Top-left panel: Fe~\texttt{I} continuum at $\Delta\lambda=+0.05$~nm. Top-right panel:
  Ca~\texttt{II} continuum at $\Delta\lambda=+0.24$~nm. Middle panels: Ca~\texttt{II} filtergrams at
   $\Delta\lambda=\pm0.03$~nm. Bottom-left panel: line-core image of H$\alpha$. Bottom-right
    panel: line-core image of Ca~\texttt{II}. The white arrow points to disk center.
    The white box encloses the big umbra: the region where we study the chromospheric spectral
    emission. Pixels \textbf{\textit{a}}, \textbf{\textit{b}} and \textbf{\textit{c}} are adopted
    to represent the two parts of the umbra and a surrounding granulation.
 }
 \label{fig1}
\end{figure}

Fig.~\ref{fig1} shows simultaneous filtergrams of these three
spectral lines. In the photosphere, observed in Fe~\texttt{I}
continuum (top-left panel), the granulation pattern surrounds the
observed irregular sunspot. The sunspot contains a big umbra and a
tail containing a few umbral cores.

%----------------------------------------------------------------------
%----------------------------------------------------------------------

A filamentary light-bridge divides the big umbra in two halves. The
Ca~\texttt{II} continuum image (top-right panel) shows similar
structures as well as bright filigree in the surrounding
granulation. At $\Delta\lambda=\pm0.03$~nm away of the expected line
center of Ca~\texttt{II} line, we can see its chromospheric
contribution (the middle-row panels). The superpenumbra is seen in
the line core images (the bottom row panels). The white box shown in
panels of Fig.~\ref{fig1} encloses the region, the big umbra, where
we study the chromospheric spectral emission in
Ca~\texttt{II}~854.2~nm.

\begin{figure} %Fig.~2
 \centerline{\includegraphics[angle=90, width=\textwidth,clip=]{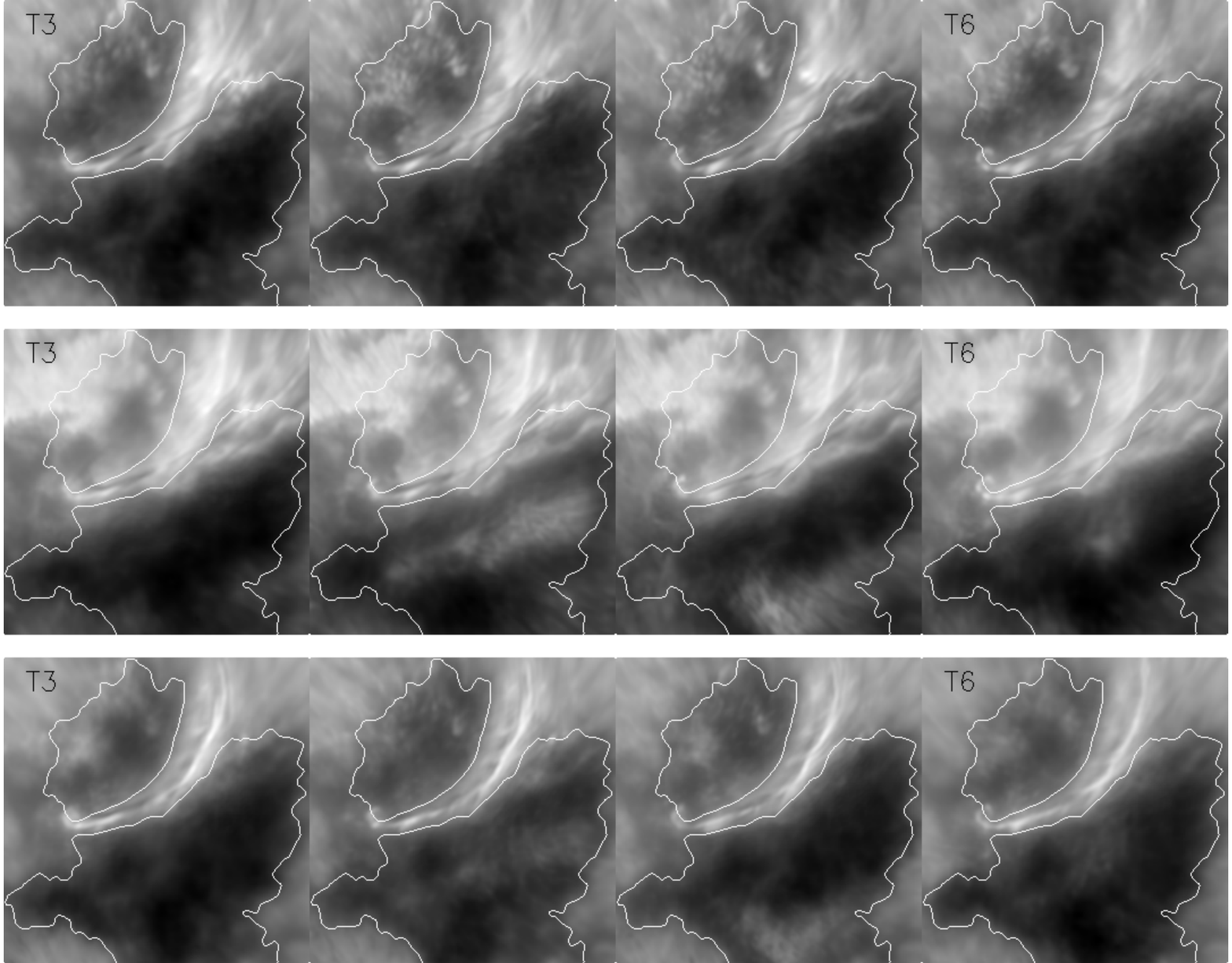}}
 \caption[]{Four successive intensity filtergrams of the Ca~\texttt{II} line, from the third
 scan through the sixth scan. Top row: near red-wing ($\Delta\lambda=+0.03$~nm) filtergrams.
 Middle row: line-core filtergrams. Bottom row: near blue-wing ($\Delta\lambda=-0.03$~nm) filtergrams.
 Time is elapsing from left to right. All images in each row were displayed with the same logarithmic
 scaling. White contours enclose the active and quiet umbra as in
 Fig.~\ref{fig1}. The size of each map is $270\times270$ pixels ($16^{\prime\prime}\times16^{\prime\prime}$)
}
 \label{camov}
\end{figure}

As can be seen in top row of Fig.~\ref{camov}, the upper half of the
big umbra shows bright dot-like structures in near-red wing
filtergrams. Near-blue wing filtergrams, bottom row of
Fig.~\ref{camov}, do not clearly show such bright dot-like
structures (smeared bright structures with lower contrast are
observed instead). When we get closer to the line core (see middle
row of Fig.~\ref{camov}), we can see the chromosphere. Then, these
dot-like structures are changing to a continuous bright structure (a
bright cloud) which is steadily seen during the whole observation so
that we could not recognize the upper half of the umbra from its
adjacent penumbra. Also, the steady bright cloud is obviously seen
in H$\alpha$ filtergrams (see bottom-left panel in Fig.~\ref{fig1}).

Running waves and umbral flashes are seen moving inside the big
umbra and in the tail of umbral cores as well and passing the
penumbra in the time series of line-core and near blue-wing
filtergrams of Ca~\texttt{II} line (see middle and bottom rows of
Fig.~\ref{camov}). Near red-wing filtergrams of Ca~\texttt{II} do
not clearly show signatures of umbral flashes.

\begin{figure} %Fig.~3
 \centerline{\includegraphics[width=0.6\textwidth,clip=]{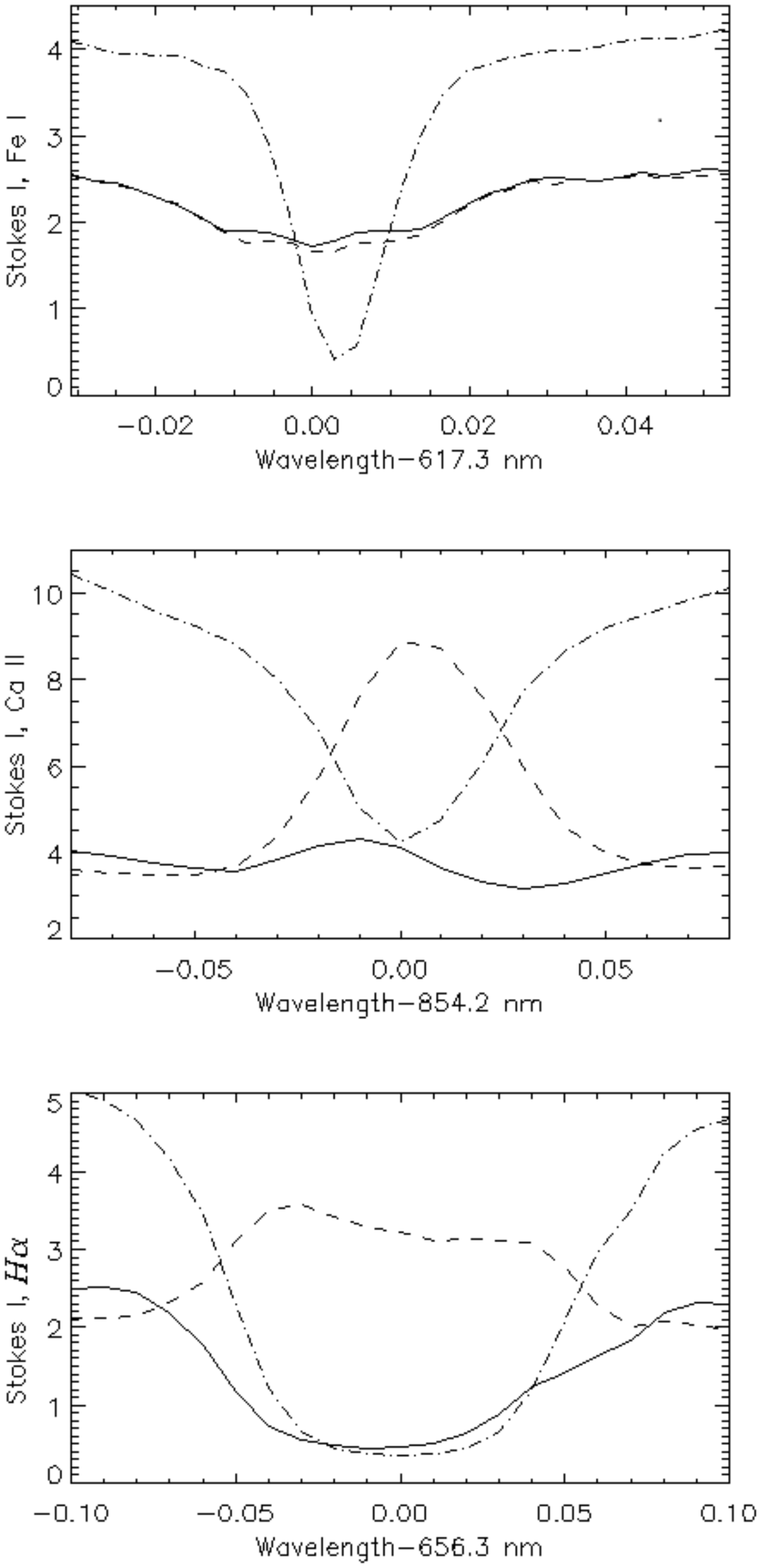}}
 \caption[]{Examples of Stokes~\textit{I} of Fe~\texttt{I} (top panel), Ca~\texttt{II} (middle panel)
 and H$\alpha$ (bottom panel) lines at the pixel \textbf{\textit{a}} in the quiet umbra (solid lines),
 pixel \textbf{\textit{b}} in the active umbra (dashed lines) and pixel \textbf{\textit{c}} in the surrounding
 granulation (dash-dotted lines). For a better visualization, the Stokes~\textit{I} of pixel \textbf{\textit{c}}
 in top panel was shifted down by 5 units and the Stokes~\textit{I} of pixel \textbf{\textit{c}} in bottom panel
 was divided by 3. All profiles were extracted from the third scan.
 }
 \label{iprof}
\end{figure}

The bright cloud seen in the chromosphere of the upper half of the
big umbra is a signature of steadily strong emissions in both
Ca~\texttt{II} and H$\alpha$ spectral lines in this part of the
umbra. Fig.~\ref{iprof} shows samples of Stokes~\textit{I} profiles
of the observed spectral lines at two pixels in the big umbra, pixel
\textbf{\textit{a}} in the lower half umbra (\textit{quiet umbra})
and pixel \textbf{\textit{b}} in the upper half umbra
(\textit{active umbra}), as well as at pixel \textbf{\textit{c}} in
the surrounding granulation. These three pixels are marked as
\textbf{\textit{a}}, \textbf{\textit{b}} and \textbf{\textit{c}},
respectively, in Fig.~\ref{fig1}.

In the quiet umbra, the Ca~\texttt{II} line shows emission signature
when umbral flashes are passing through the umbra while the
H$\alpha$ line shows a blue-shifted absorption profile. However,
active umbra shows steady and strong spectral emissions in both
Ca~\texttt{II} and H$\alpha$ lines, irrespective of evolving umbral
flashes. Nevertheless, running waves and umbral flashes only
slightly and temporarily change the brightness of the bright cloud
in the line-core filtergrams (see middle row of Fig.~\ref{camov})
which is essentially because of the effect of the Doppler shift of
the emission part of the line profile. Also, umbral flashes and the
resulting Doppler-shifted emission part of the profiles can change
more conspicuously the shape and brightness of the bright cloud at
the near-wing filtergrams (see top and bottom rows in
Fig.~\ref{camov}).

\begin{figure} %Fig.~4
 \centerline{\includegraphics[angle=90, width=0.95\textwidth,clip=]{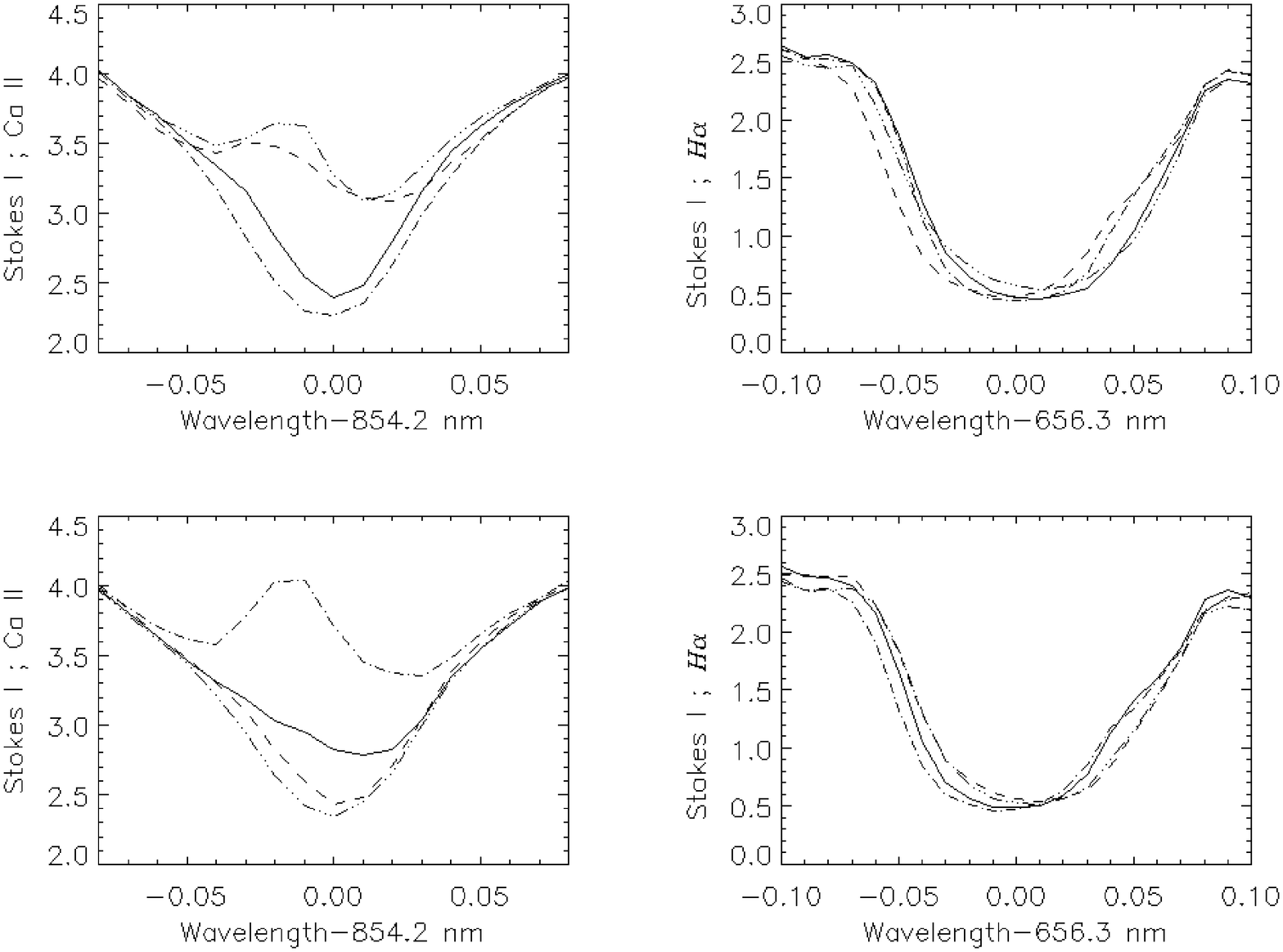}}
 %{IV_vs_r_0_48allpixels.eps}}
 \caption[]{First row: four successive Stokes~\textit{I} profiles of Ca~\texttt{II} (left panel)
 and H$\alpha$ (right panel) for pixel \textbf{\textit{a}} in quiet umbra. Second row: the next four successive
 Stokes~\textit{I} profiles the same as the first row. Time is elapsing through solid, dashed,
 dash-dotted and dash-triple-dotted lines.
 }
 \label{istokes}
\end{figure}

As can be seen in Fig.~\ref{iprof}, the Stokes~\textit{I} profile of
the Ca~\texttt{II} line for pixel \textbf{\textit{a}} shows a
conspicuous emission in the blue wing, while the Stokes~\textit{I}
profile of H$\alpha$ line shows a blue-shift and a deformation in
its red wing. Also, this deformation is seen in the red wing of
Stokes~\textit{I} of pixel \textbf{\textit{c}} (see bottom panel of
Fig.~\ref{iprof}). To have a detailed comparison, Fig.~\ref{istokes}
displays eight successive Stokes~\textit{I} profiles of
Ca~\texttt{II} and H$\alpha$  for pixel \textbf{\textit{a}}. We can
see a correlation between the occurrence of emission in the
Ca~\texttt{II} line and a global blue-shift in the H$\alpha$ profile
along with a deformation in its red wing. There are two exceptions
in the first row panels for dash-dotted and dash-triple dotted
lines: They show a time delay between emission signatures in the
Ca~\texttt{II} line and the global blue shift in H$\alpha$ line
maybe because of the time difference between the Ca~\texttt{II} and
H$\alpha$ scans. The observed deformation in the red wing of the
Stokes~\textit{I} of H$\alpha$ line can be a signature of velocity
gradients connected with the oscillations. The inequality of the
blue and red continua of the H$\alpha$ line is probably caused by
weak absorption lines of Co~\texttt{I} and atmospheric H$_{2}$O in
the region 656.34 - 656.42~nm.

%----------------------------------------------------------------------
%----------------------------------------------------------------------
%----------------------------------------------------------------------
%----------------------------------------------------------------------
%----------------------------------------------------------------------
\begin{figure} %Fig.~5
 \centerline{\includegraphics[angle=90, width=\textwidth,clip=]{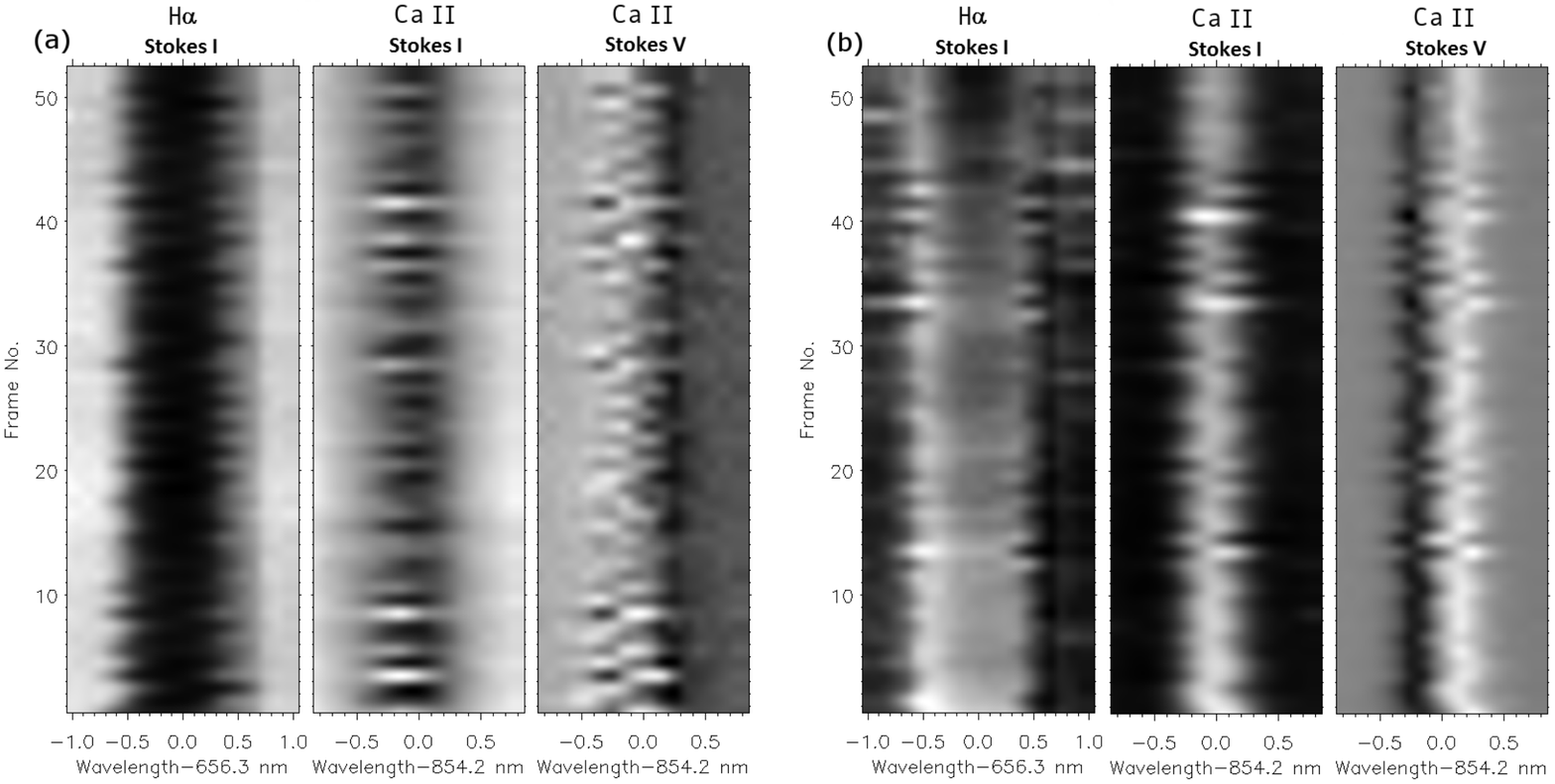}}
 \caption[]{Temporal changes of H$\alpha$ (Stokes~\textit{I}) and Ca~\texttt{II} (Stokes~\textit{I}
 and \textit{V}) in pixel \textbf{\textit{a}} (left panel) and pixel \textbf{\textit{b}} (right panel).
 }
 \label{lambda_t}
\end{figure}
%----------------------------------------------------------------------

Fig.~\ref{lambda_t} illustrates temporal changes ($\lambda$-t
diagrams) of H$\alpha$ Stokes~\textit{I} and Ca~\texttt{II}
(Stokes~\textit{I} and \textit{V}). The Ca~\texttt{II} line shows
discrete emissions, corresponding to umbral flashes, in pixel
\textbf{\textit{a}} (quiet umbra, left panel) and a steady emission
in pixel \textbf{\textit{b}} (active umbra, right panel), where
umbral flashes are also seen. The ``polarity reversal'' (a reversed
sign not a true polarity reversal) in Ca~\texttt{II}
Stokes~\textit{V} occurs when a strong emission appears. Pixel
\textbf{\textit{b}} shows a steady reversed polarity sign with
time-varying Doppler shifts. In the quiet umbra, the H$\alpha$ line
has an absorption profile while in the active umbra, H$\alpha$ is
steadily in emission. Both the absorption and emission profiles show
blue-shifts and asymmetries during umbral flashes.

\section{The Cloud Model}\label{bcm}
The aim of this paper is to study Doppler velocities in the
chromosphere above the umbra where we observe emission signatures in
the Ca~\texttt{II} line.

Spectral inversion techniques based on cloud model \cite{tzio07} are
extremely useful for the study of properties and dynamics of various
chromospheric cloud-like structures.

Cloud models refer to models describing the transfer of radiation
through structures located above the solar photosphere. Such
cloud-like structures seem to absorb the incident radiation coming
from below and to add some emissions. The mentioned absorption and
emission processes are described by the formal solution of the
radiative transfer equation
\begin{equation}
I_{\lambda}=I_{0\lambda}e^{-\tau_{\lambda}}+\int_{0}^{\tau_{\lambda}}S_{t}e^{-t_{\lambda}}dt_{\lambda}
\end{equation}
where $I_{\lambda}$ is the observed intensity, $I_{0\lambda}$ is the
incident radiation to the cloud from below, $\tau_{\lambda}$ is the
optical thickness and $S_{t}$ is the source function which can be a
function of optical depth inside the cloud. The first term of the
right hand part of the Eq. (1) represents the absorption of the
incident radiation by the cloud while the second term represents the
added emission by the cloud itself.

\subsection{Thin Cloud Model}\label{tcm}
We follow the Beckers cloud model \cite{beckers64} by assuming that
\textbf{a)} the source function and the LOS velocity are constant
through the whole cloud, \textbf{b)} the optical thickness of the
cloud has a Gaussian wavelength dependence with a constant Doppler
width, and more important, \textbf{c)} the cloud is optically thin
to produce a strong emission superimposed on an absorption profile.
Heinzel and Schmieder \cite{heinzel94}, using a grid of many non-LTE
models of prominence-like structures, found that a constant source
function corresponds to a low-pressure, optically thin structure.
%----------------------------------------------------------------------
\begin{figure} %Fig.~ 6
 \centerline{\includegraphics[angle=90, width=\textwidth,clip=]{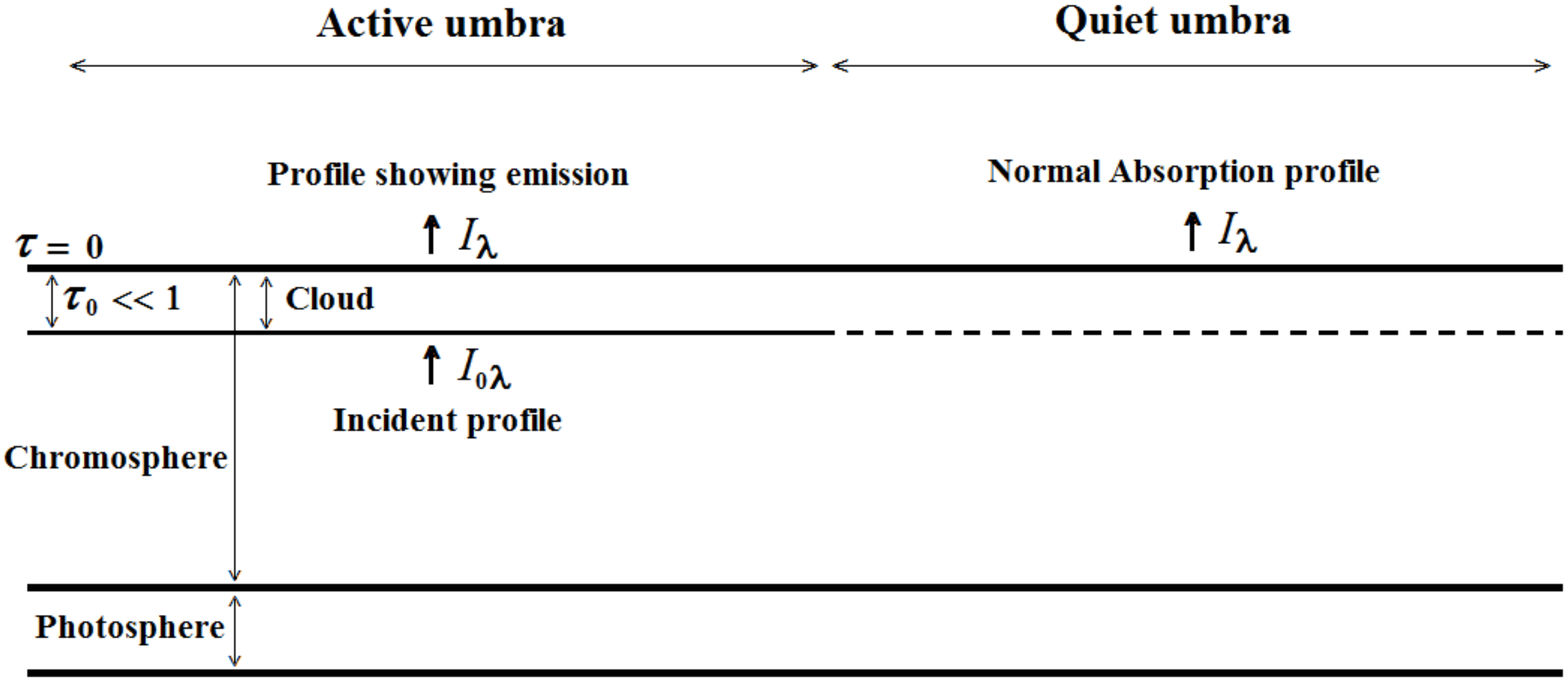}}
 \caption[]{Geometry of the chromospheric cloud. The source function is always non-zero
 in active umbra but, in quiet umbra the source function has a non-zero value when an
 umbral flash is propagating or evolving.
 }
 \label{cloud_geo}
\end{figure}
%----------------------------------------------------------------------
Fig.~\ref{cloud_geo} shows the geometry of the cloud in the solar
atmosphere; the cloud can be defined as upper layers of umbral
chromosphere that are optically thin. There must be an active
mechanism inside these layers producing a strong emission.

By these considerations, Eq.~(1) is reduced to
\begin{equation}
I_{\lambda}=I_{0\lambda}-\tau_{\lambda}I_{0\lambda}+\tau_{\lambda}S
\end{equation}
The third term on the right hand side of Eq. (2)
$E_{\lambda}=\tau_{\lambda}S$ is the added total emission term that
can be interpreted as the wavelength-dependent total emission of the
cloud. This means that we do expect that a profile showing emission
is described by the product $\tau_{\lambda}S$, not by $S$ and
$\tau_{\lambda}$ separately. Eq.~(2) can be rewritten as
\begin{equation}
I_{\lambda}=I_{0\lambda}+\tau_{\lambda}(S-I_{0\lambda})
\end{equation}
The optical thickness of the cloud is described by a Gaussian
function as
\begin{equation}
\tau_{\lambda}=\tau_{0}e^{-(\frac{\lambda-\lambda_{P}}{W})^{2}}
\end{equation}
where  $\tau_{0}$($<<1$: cloud is optically thin) is the peak
optical thickness, $W$ is the Doppler width and $\lambda_{P}$ is the
Doppler shifted wavelength of the peak absorption which is related
to the LOS velocity $V_{C}$ of the cloud and the
reference-line-center wavelength $\lambda_{0}$ via

\begin{equation}
V_{C}=\frac{\lambda_{P}-\lambda_{0}}{\lambda_{0}}c
\end{equation}

In this relation, $c$ is the speed of light. By the definition of
$\tau_{\lambda}$ in Eq. (4), the wavelength-dependent total emission
of the cloud $E_{\lambda}$ has the same Gaussian function as
$\tau_{\lambda}$ with a peak total emission of $E_{0}=\tau_{0}S$.

The Doppler width $W$ depends on temperature $T$ and micro-turbulent
velocity $\xi$ through the relationship
\begin{equation}
W=\frac{\lambda_{0}}{c}\sqrt{\frac{2k_{B}T}{m}+\xi^{2}}
\end{equation}
where $m$ is the atomic mass of the absorbing/emitting element and
$k_{B}$ is the Boltzmann constant.

The four adjustable/free parameters of the thin cloud model are the
source function $S$, the Doppler width $W$, the peak optical
thickness $\tau_{0}$ and the LOS velocity $V_{C}$. All these
parameters are assumed to be constant through the cloud structure
that is responsible for the observed emission.

As mentioned before, in this work we are only interested in
obtaining the Doppler velocity map in the selected big umbra (see
Fig.~\ref{fig1}) by applying the described inversion based on cloud
model on Stokes~\textit{I} profiles of Ca~\texttt{II}~854.2~nm. As
far as the LOS velocity values are concerned, the cloud model can be
used since velocity is the most model-independent parameter.

\subsection{Background Incident Profile}\label{bgprof}
The profile of background incident light on the cloud from below is
defined as the average of normal absorption profiles in quiet umbra
with the same continuum/far-wing intensity as the observed profile.
This assumption is based on the fact that pixels in quiet umbra
showing emission (such as pixel \textbf{\textit{a}} marked in
Fig.~\ref{fig1}) show small fluctuations at their continuum/far-wing
intensity during the whole observation (see the left panels of
Fig.~\ref{istokes}). Also, this fact can be seen in Figs. 4 \& 5 in
de la Cruz Rodriguez et al. \cite{dela13aa} who have studied umbral
flashes in a sunspot chromosphere. Also, Fig.~\ref{fig1} in de la
Cruz Rodriguez et al. \cite{dela13apj} who have studied a magnetic
chromospheric region in quiet Sun, displays that the photospheric
wings of ``reversed-core'' profiles of Ca~\texttt{II} line at
854.2~nm (showing emission) are identical in brightness and shape to
those of quiet region profiles that show a typical absorption shape.

%----------------------------------------------------------------------
\begin{figure} %Fig.~7
 \centerline{\includegraphics[angle=90, width=0.7\textwidth,clip=]{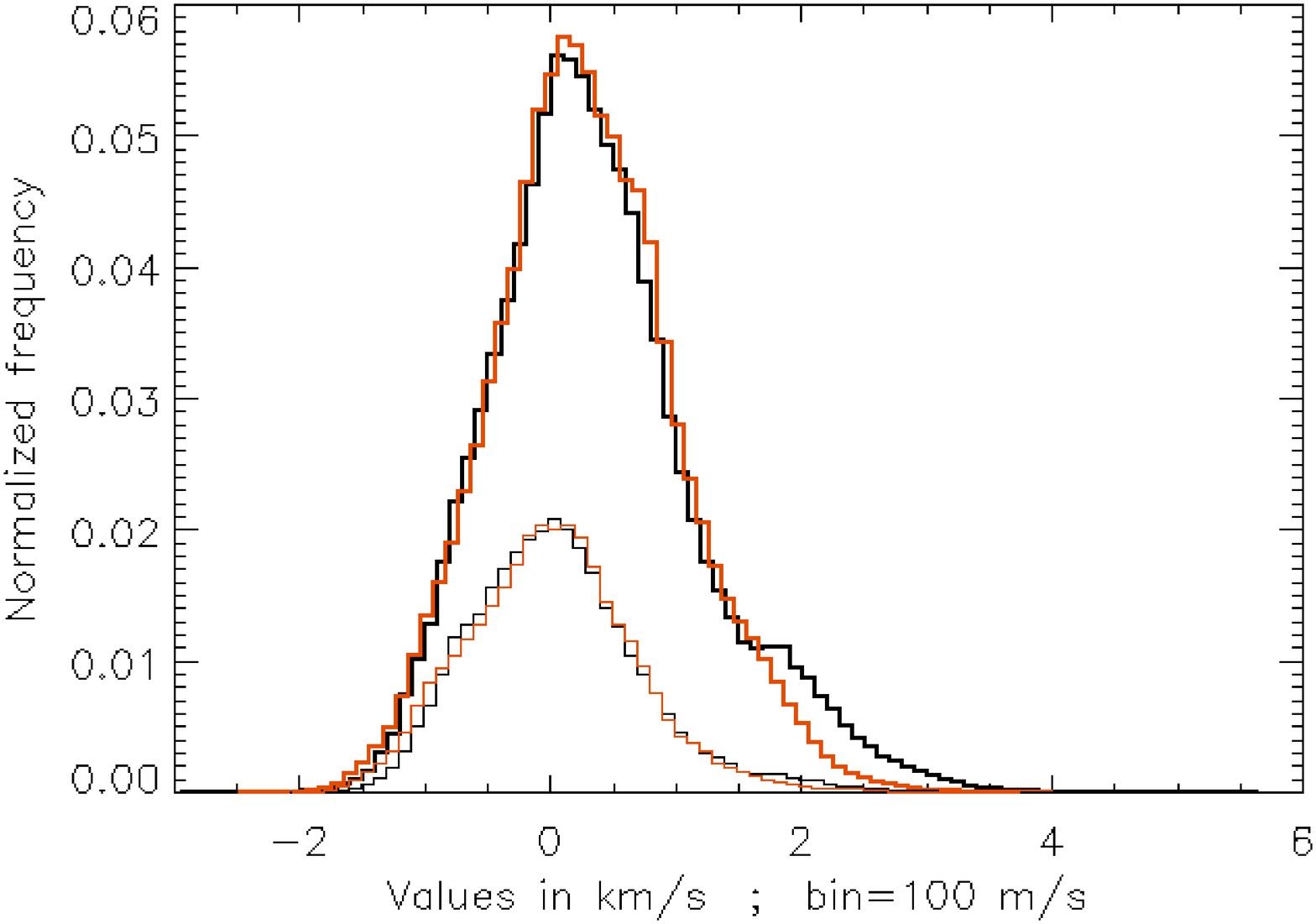}}
 \caption[]{The histogram of core velocities (black lines) and center-of-gravity velocities (red lines)
 of normal absorption profiles obtained in the whole studied area (thick lines) and in the umbra (thin lines).
 }
 \label{hist_vbg7}
\end{figure}
%----------------------------------------------------------------------

By ignoring the small absorption through the upper layers of the
quiet umbral chromosphere that are optically thin (see Eq. (2)),
this averaged profile can be the best sample profile describing the
incident profile into the cloud.

In this work, a normal absorption profile is selected by finding the
curvature of the Stokes~\textit{I} profile around the expected
position of line center; if two sets of four adjacent wavelength
positions (among seven wavelength positions around the line-core)
show a positive concavity, the profile can be a normal absorption
profile. For more certainty, some considerations based on the
equivalent line width were applied; according to the standard
definition of equivalent line width, the profiles showing emission
have smaller line width with respect to that of normal absorption
profiles. In the case of strong emission, we obtain a negative line
width.

Fig.~\ref{hist_vbg7} displays the histogram of line-core velocities
using a parabola fit (thick black line) and center-of-gravity
velocities - between the intensity level of line-core and 60\% of
line depth - (thick red line) for all obtained normal absorption
profiles in the studied umbra. Thin solid lines in
Fig.~\ref{hist_vbg7} show the velocity distributions of umbral
profiles excluding the light bridge and penumbral parts in the
studied area. As it is obvious from Fig.~\ref{hist_vbg7}, we have
adopted the peak position of the distribution of core velocities in
umbra as the zero-velocity reference $\lambda_{0}$.

To compute the background incident profile we make the average of 50
(at most) normal absorption profiles with the same far-wing
intensity as the observed profile. To prevent an artificial line
broadening, we select only normal profiles whose line-core
velocities are within $\pm700~\textrm{m}~\textrm{s}^{-1}$
(equivalent to line-core positions within
$\Delta\lambda=\pm0.02$~nm). Then, we shift these profiles to
zero-velocity reference and compute the averaged profile.

\subsection{Model Fitting}\label{modelfit} Using Eq. (4), Eq. (3) can
be rewritten as
\begin{equation}
I_{\lambda}-I_{0\lambda}=\tau_{0}e^{-(\frac{\lambda-\lambda_{P}}{W})^{2}}(S-I_{0\lambda})
\end{equation}
We use an iterative method that consists of three steps.

1) First, estimate of free parameters: at the first step, we ignore
$I_{0\lambda}$ on the right hand side of Eq. (7). This assumption is
reasonable since $\tau_{0}$ is much less than unity. Then the value
of $S-I_{0\lambda}$ on the right hand side must be large enough to
adjust the difference value on the left hand side, although the
position of $\lambda_{P}$ and the value of $W$ have important roles
in this adjustment. Then, Eq.~(7) is reduced to
\begin{equation}
I_{\lambda}-I_{0\lambda}=S\tau_{0}e^{-({\frac{\lambda-\lambda_{P}}{W}})^{2}}
\end{equation}
By fitting a Gaussian function to the difference profile
$I_{\lambda}-I_{0\lambda}$, we can calculate the first estimate of
two parameters $\lambda_{P}$ and $W$, and an estimate for the peak
total emission $S\tau_{0}$. Assuming an arbitrary small value for
$\tau_{0}$ (e.g. $\tau_{0}=0.05$) as an initial value, we obtain the
first estimate of the source function $S$.

2) If we substitute the first estimate of $S$ in Eq.~(7), the left
hand side of the following equation obeys a Gaussian function as
\begin{equation}
\frac{I_{\lambda}-I_{0\lambda}}{S-I_{0\lambda}}=\tau_{0}e^{-(\frac{\lambda-\lambda_{P}}{W})^{2}}
\end{equation}
In this step, by finding the best fit of a Gaussian function on the
difference-ratio profile (the left hand side of Eq.~(9)), we
calculate new values for the three free parameters  $\tau_{0}$,
$\lambda_{P}$ and $W$. Then, at $\lambda=\lambda_{P}$, the new value
of $S$ can be calculated from the fitted difference-ratio profile.
In this step, we calculate the sum of the squared differences
between the observed Stokes~\textit{I} profile and the fitted one to
compare the quality of the fit with the quality of the next fit.

3)  Substituting the new value of $S$ in Eq.~(9), we can repeat the
procedure of step (2) to obtain the best new values for all free
parameters. The iteration can be repeated till we obtain the minimum
value of the squared differences or the relative change of squared
differences reaches less than 0.1\%. This procedure is robust and in
most cases, after at most five iterations, we will obtain the best
fitted values.

Since the core position of the incident profile can affect the
obtained LOS cloud velocity $V_{C}$, we have repeated the described
iteration method using incident profiles shifted by core velocities
up to 5.5~km~s$^{-1}$, both red- and blue-shift. Then, we selected
the four free parameters for the best final fit. Thereby, we
introduced a new free parameter, the core velocity of the incident
background profile $v_{bg}$. By assuming $\tau_{0}=0.05$ as an
initial value for the peak optical thickness of the cloud, in the
followings we describe the results of the iteration method.

\begin{figure} %Fig.~8
 \centerline{\includegraphics[angle=90, width=\textwidth,clip=]{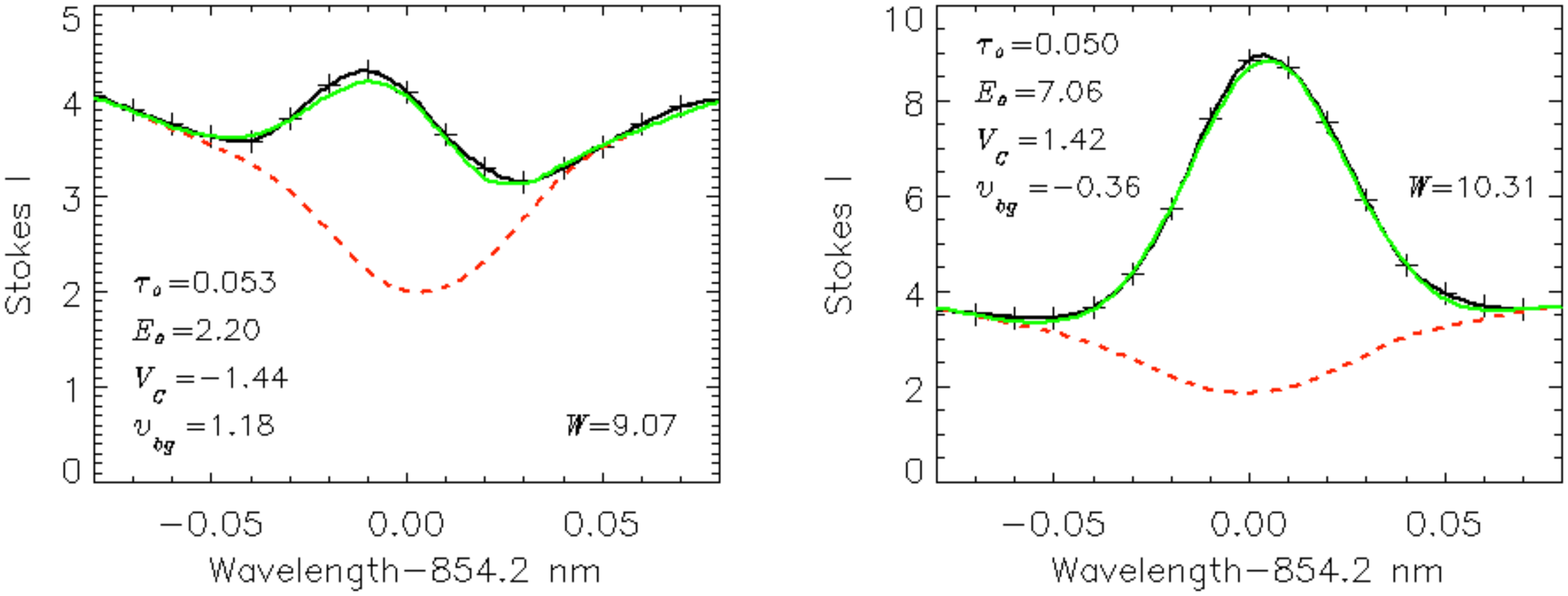}}
 \caption[]{Examples of the observed (plus symbols and black solid line), the background profile
 (red dashed line) and the final fitted profile (green solid line) for two selected sample profiles:
 in quiet umbra (left panel; pixel \textbf{\textit{a}}) and in active umbra (right panel; pixel \textbf{\textit{b}}).
 The best fitted values of $\tau_{0}$, $E_{0}$, $V_{C}$, $v_{bg}$ and $W$ are shown in each panel.
 }
 \label{fit_ab}
\end{figure}

Fig.~\ref{fit_ab} shows two examples of the observed profile (black
solid line), the background profile (red dashed line) and the final
fitted profile (green solid line) for the two sample pixels
\textbf{\textit{a}} in the quiet umbra (left panel) and
\textbf{\textit{b}} in the active umbra (right panel).

%----------------------------------------------------------------------
%----------------------------------------------------------------------

\section{Results}\label{res}
\subsection{Doppler Velocities and Line Widths}\label{dopvel}
\begin{figure} %Fig.~9
 \centerline{\includegraphics[angle=90, width=0.9\textwidth,clip=]{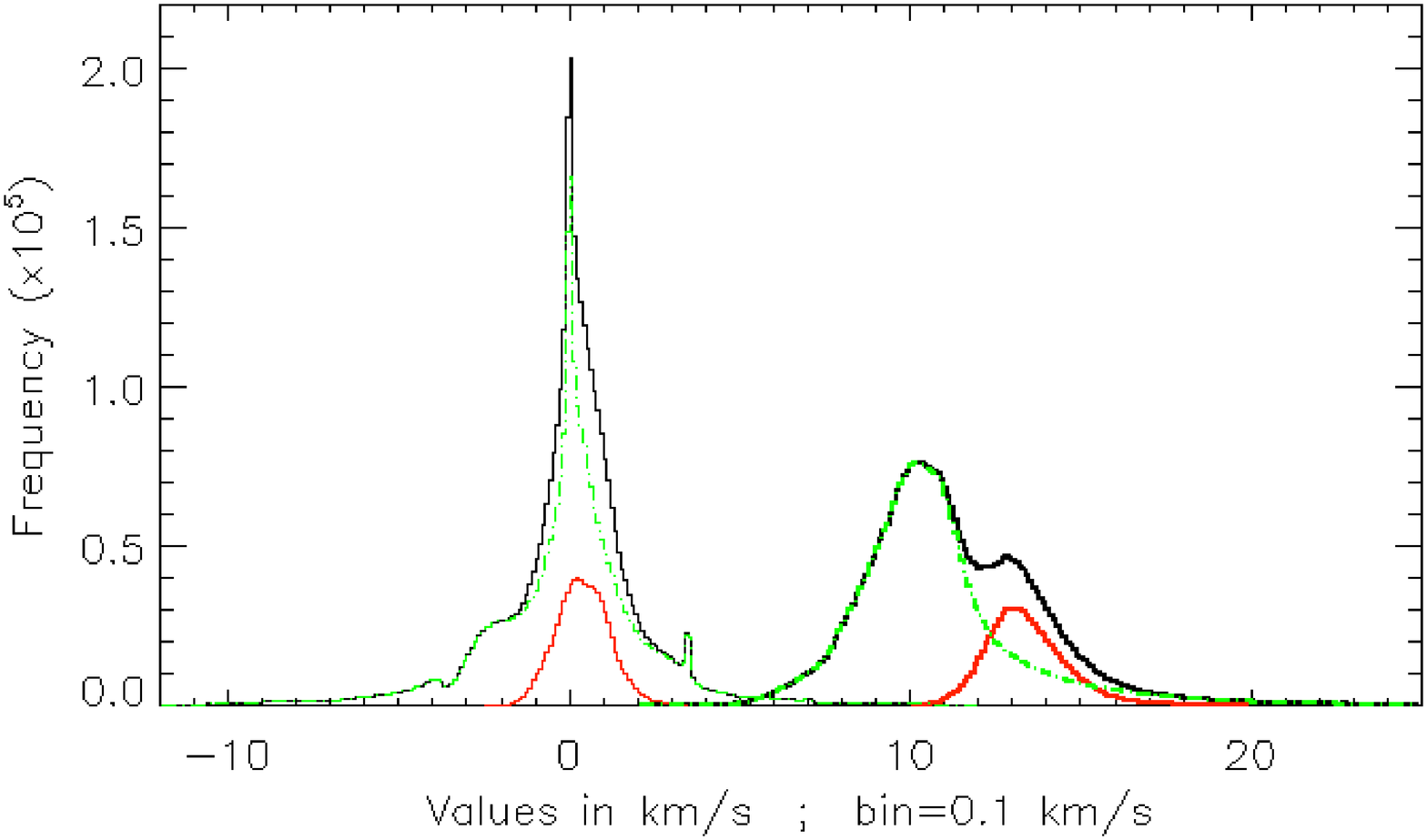}}
 \caption[]{Histograms of the Doppler velocities (thin lines) and Doppler widths (thick lines)
 during the whole observation. Black solid lines, red solid lines and green dash-dotted lines
 display the corresponding histograms for all pixels, pixels that do not show emission signature
 and those showing an emission, respectively. Negative Doppler velocities correspond to upflows.
 }
 \label{hist_velw}
\end{figure}

The thin cloud model described is applied only on profiles showing a
spectral emission signature to obtain the LOS velocity $V_{C}$ of
the cloud (see Eq.~(5)). However, the center-of-gravity method
(between the intensity level of line-core and 60\% of line depth) is
applied to the profiles keeping their absorption shape, especially
in the quiet umbra, to obtain an averaged Doppler velocity along the
formation height of the spectral line core. Fig.~\ref{hist_velw}
shows the histogram of the obtained Doppler velocities (thin lines)
for all pixels, pixels without the emission signature, and those
showing the emission, and the histogram of the Doppler widths (thick
lines) during the whole observation.

The Doppler width of a profile without emission signature is
obtained using a Gaussian fit. Black solid lines, red solid lines
and green dash-dotted lines in Fig.~\ref{hist_velw}, display the
corresponding histograms for all pixels, pixels that do not show
emission signature and those showing an emission, respectively.

\begin{figure} %Fig.~10
 \centerline{\includegraphics[angle=90, width=\textwidth,clip=]{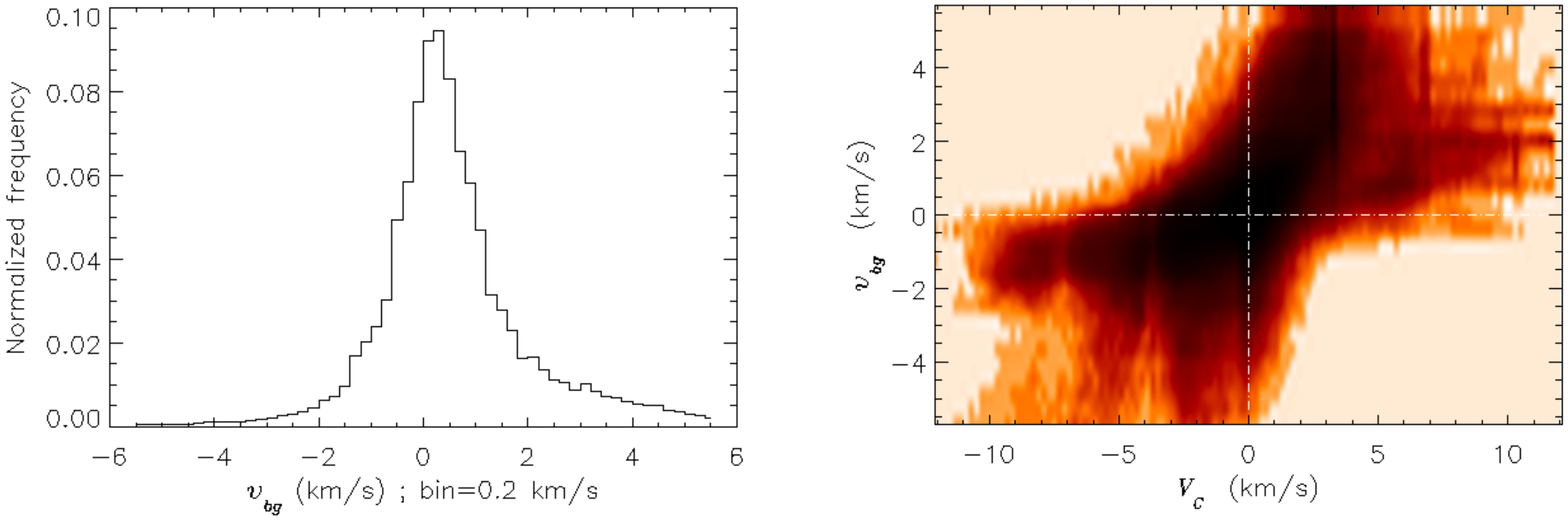}}
 \caption[]{Left panel: the histogram of $v_{bg}$. Right panel: scatter plot of $v_{bg}$
 versus $V_{C}$. The color map scales the population/frequency of each point; darker
 region means more frequency.
 }
 \label{hist_vbg10}
\end{figure}

The histogram of $v_{bg}$, the line-core velocity of the incident
background profiles after finding the best fit at the end of the
iteration process, is plotted in the left panel of
Fig.~\ref{hist_vbg10}. The asymmetry of this histogram is similar to
the asymmetry of graphs in Fig.~\ref{hist_vbg7} and the
thin-red-line graph in Fig.~\ref{hist_velw}, which supports the
correctness of our iteration procedure to obtain the best fitted
parameters. The relation between $v_{bg}$ and $V_{C}$ is illustrated
in the right panel of Fig.~\ref{hist_vbg10}. A possible correlation
between $v_{bg}$ and $V_{C}$ is seen in this plot. In many pixels,
the obtained velocities reflect the propagation of waves from the
photosphere to the cloud so that the photospheric and chromospheric
velocities can be partially correlated.

\begin{figure} %Fig.~11
 \centerline{\includegraphics[angle=90, width=\textwidth,clip=]{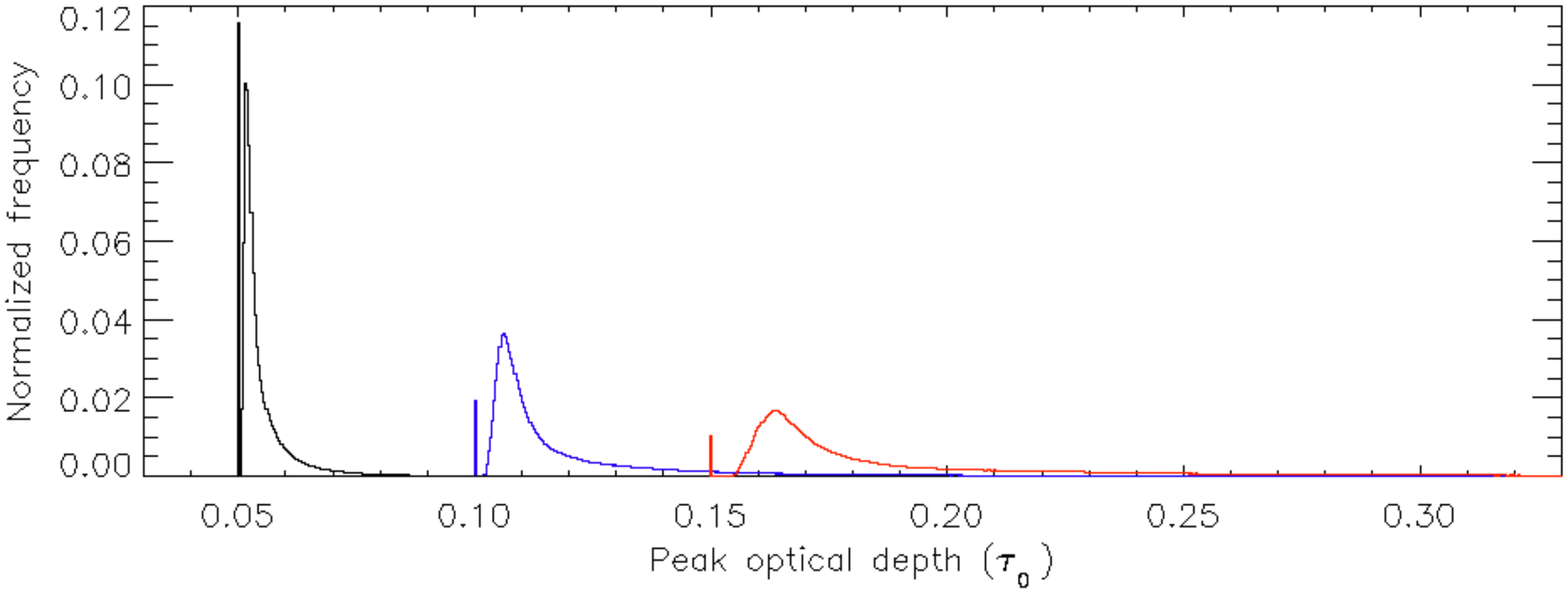}}
 \caption[]{Histograms of the obtained peak optical thicknesses for three different
 initial values of $\tau_{0}$: 0.05 (black line), 0.1 (blue line) and 0.15 (red line).
 }
 \label{hist_tau0}
\end{figure}

About 0.4\% of pixels show peak optical thickness $\tau_{0}>0.1$.
Fig.~\ref{hist_tau0} displays the histogram of the fitted $\tau_{0}$
values (black solid line).

%----------------------------------------------------------------------
%----------------------------------------------------------------------
%----------------------------------------------------------------------

\subsection{Influence of the Optical Thickness on the
Results}\label{tau0effect} 
\begin{figure} %Fig.~12
 \centerline{\includegraphics[angle=90, width=0.7\textwidth,clip=]{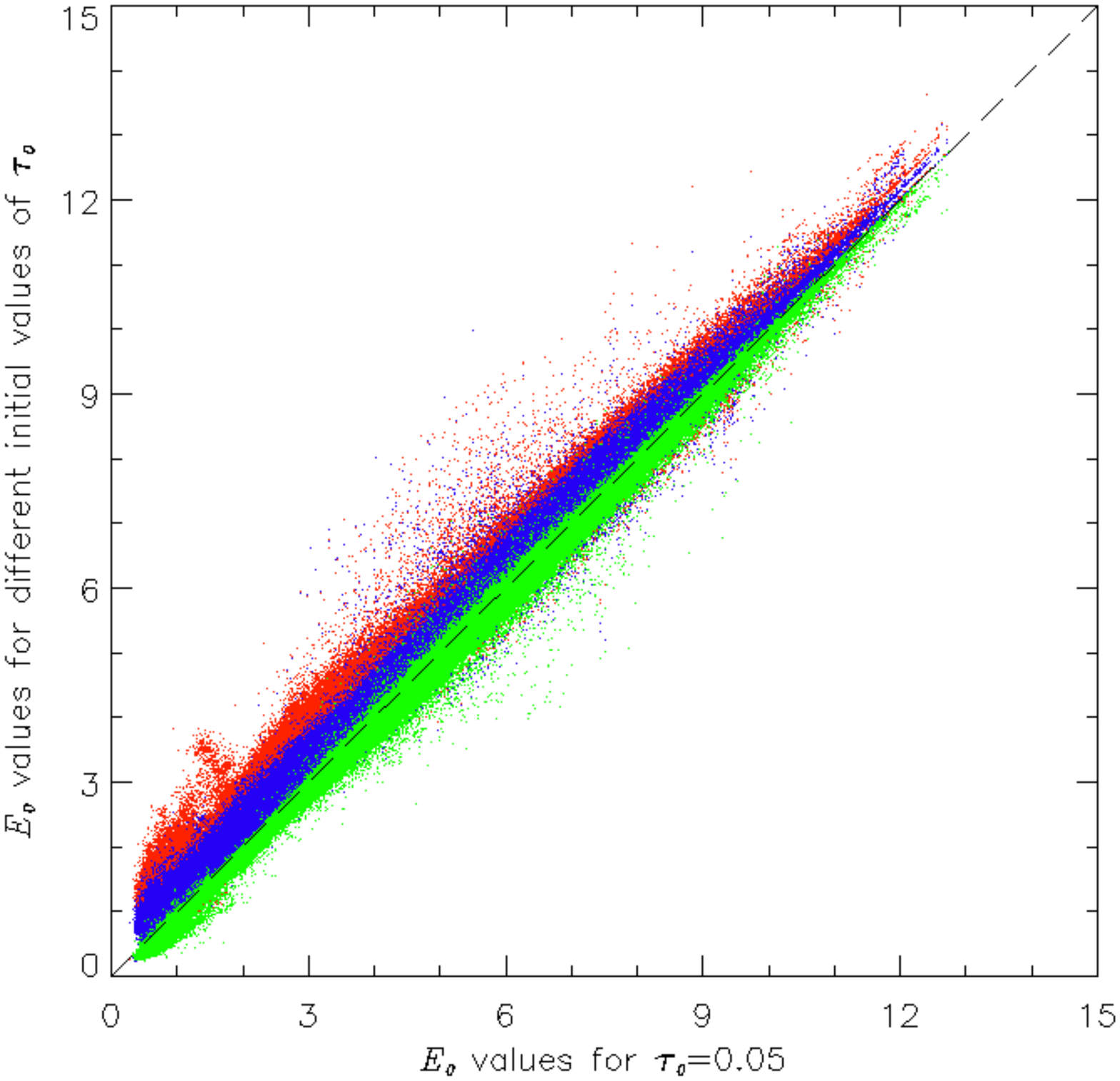}}
 \caption[]{Scatter plots of peak total emission ($E_{0}=S\tau_{0}$) for different
 initial values of $\tau_{0}$ (0.1: red dots; 0.15: blue dots; 0.02: green dots)
 versus the one-to-one correspondence of $E_{0}$ for the initial value $\tau_{0}=0.05$.
 }
 \label{scater_e0}
\end{figure}

\begin{figure} %Fig.~13
 \centerline{\includegraphics[angle=90, width=0.7\textwidth,clip=]{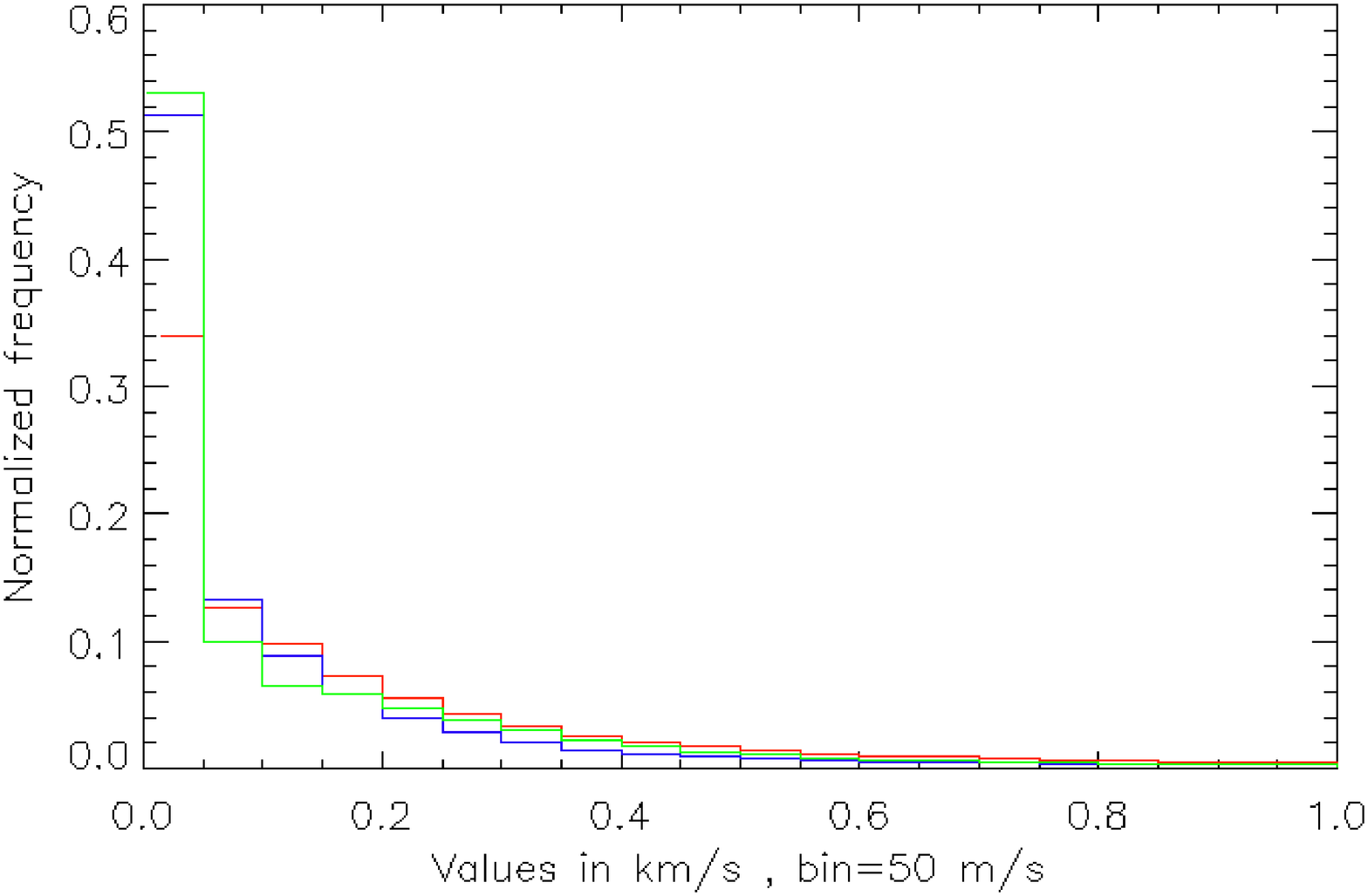}}
 \caption[]{Histograms of the absolute values of differences between the obtained Doppler
 velocities for two different initial values of peak optical thickness: red line for
 $\tau_{0}=0.05$ and $\tau_{0}=0.15$; blue line for $\tau_{0}=0.05$ and $\tau_{0}=0.1$;
 green line for $\tau_{0}=0.05$ and $\tau_{0}=0.02$.
 }
 \label{hist_dv}
\end{figure}

If we change the initial value of
$\tau_{0}$ to a larger value 0.1 or 0.15, we will obtain similar
histograms for final fitted $\tau_{0}$ as shown in
Fig.~\ref{hist_tau0} (blue and red lines, respectively) with shifted
peaks to 0.1 and 0.15, respectively, in expense of obtaining smaller
source functions. However, in the case of $\tau_{0}<0.1$, values of
the product $S\tau_{0}$  (peak total emission, $E_{0}$) for each
pixel are almost the same. Fig.~\ref{scater_e0} shows the scatter
plots of peak total emission ($E_{0}$) for different initial values
of $\tau_{0}$ versus the one-to-one correspondence of $E_{0}$ for
the initial value $\tau_{0}=0.05$. In Figs.~\ref{scater_e0} \&
\ref{hist_dv}, we have added the results of the iteration process
for the initial value $\tau_{0}=0.02$ to have a better comparison.

Fig.~\ref{hist_dv} displays the distribution of the absolute values
of the differences between the obtained Doppler velocities for three
sets of results of $\tau_{0}=0.05$ and $\tau_{0}=0.15$ (red line),
of $\tau_{0}=0.05$ and $\tau_{0}=0.1$ (blue line) and of
$\tau_{0}=0.05$ and $\tau_{0}=0.02$ (green line). The histograms
have peaks around zero and quickly decrease towards larger values.

Both Figs.~\ref{scater_e0} \& \ref{hist_dv} demonstrate that there
is a good solution for the radiative transfer equation based on thin
cloud model (with $\tau_{0}<0.1$) for profiles showing emission. As
mentioned before, the cloud Doppler velocity and its total emission
are two important free parameters of the model that are practically
independent of the initial value of $\tau_{0}$.

\begin{figure} %Fig.~14
 \centerline{\includegraphics[angle=90, width=\textwidth,clip=]{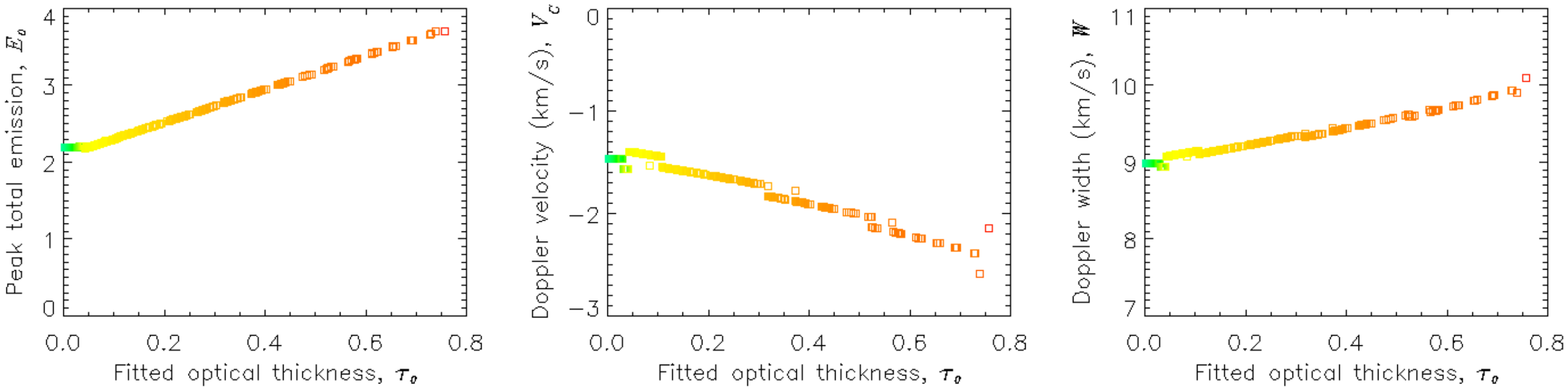}}
 \caption[]{Changes of the fitted free parameters of
 thin cloud model $E_{0}$, $V_{C}$ and $W$ versus the final different fitted $\tau_{0}$ for the
 pixel \textbf{\textit{a}} in the quiet umbra. The colored squares scale the squared differences value of
 the fitted profile and of the observed profile from minimum value (the best fit, green)
 to maximum value (the worst fit, red).
 }
 \label{fitquality_a}
\end{figure}

\begin{figure} %Fig.~15
 \centerline{\includegraphics[angle=90, width=\textwidth,clip=]{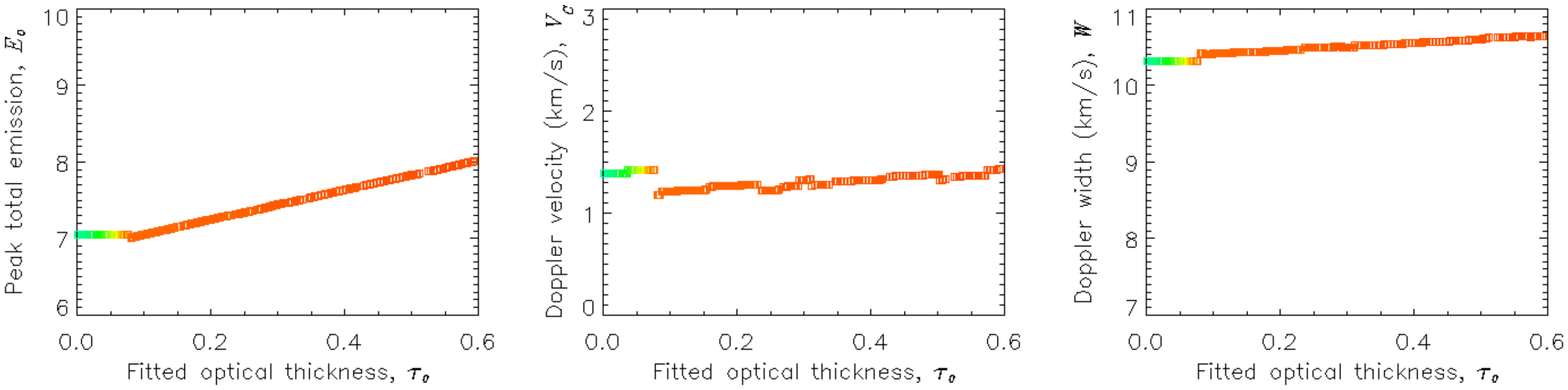}}
 \caption[]{Same as in Fig.~\ref{fitquality_a} but for the pixel \textbf{\textit{b}} in the active umbra.
 }
 \label{fitquality_b}
\end{figure}

For a deep investigation of the effect of the initial value of
$\tau_{0}$ on other free parameters of the thin cloud model, we
again consider the two pixels \textbf{\textit{a}} and
\textbf{\textit{b}} whose spectral profiles have been shown in
Fig.~\ref{fit_ab}. For each one of these pixels, we change the
initial value of $\tau_{0}$ from 0.002 to 0.5 and compare the fitted
free parameters. Figs.~\ref{fitquality_a} \& \ref{fitquality_b} show
the changes of the fitted free parameters of the model $E_{0}$,
$V_{C}$ and $W$ versus the final fitted for pixels
\textbf{\textit{a}} and \textbf{\textit{b}}, respectively. As seen
from these figures, assuming $\tau_{0}<0.1$ as an initial value for
optical thickness of the cloud, the thin cloud model retrieves
unique values for other free parameters of the model.

%----------------------------------------------------------------------
%----------------------------------------------------------------------
%----------------------------------------------------------------------
%----------------------------------------------------------------------
\subsection{Examples of Line-profile Fits}\label{fitexampl} 
\begin{figure} %Fig.~16
 \centerline{\includegraphics[width=\textwidth,clip=]{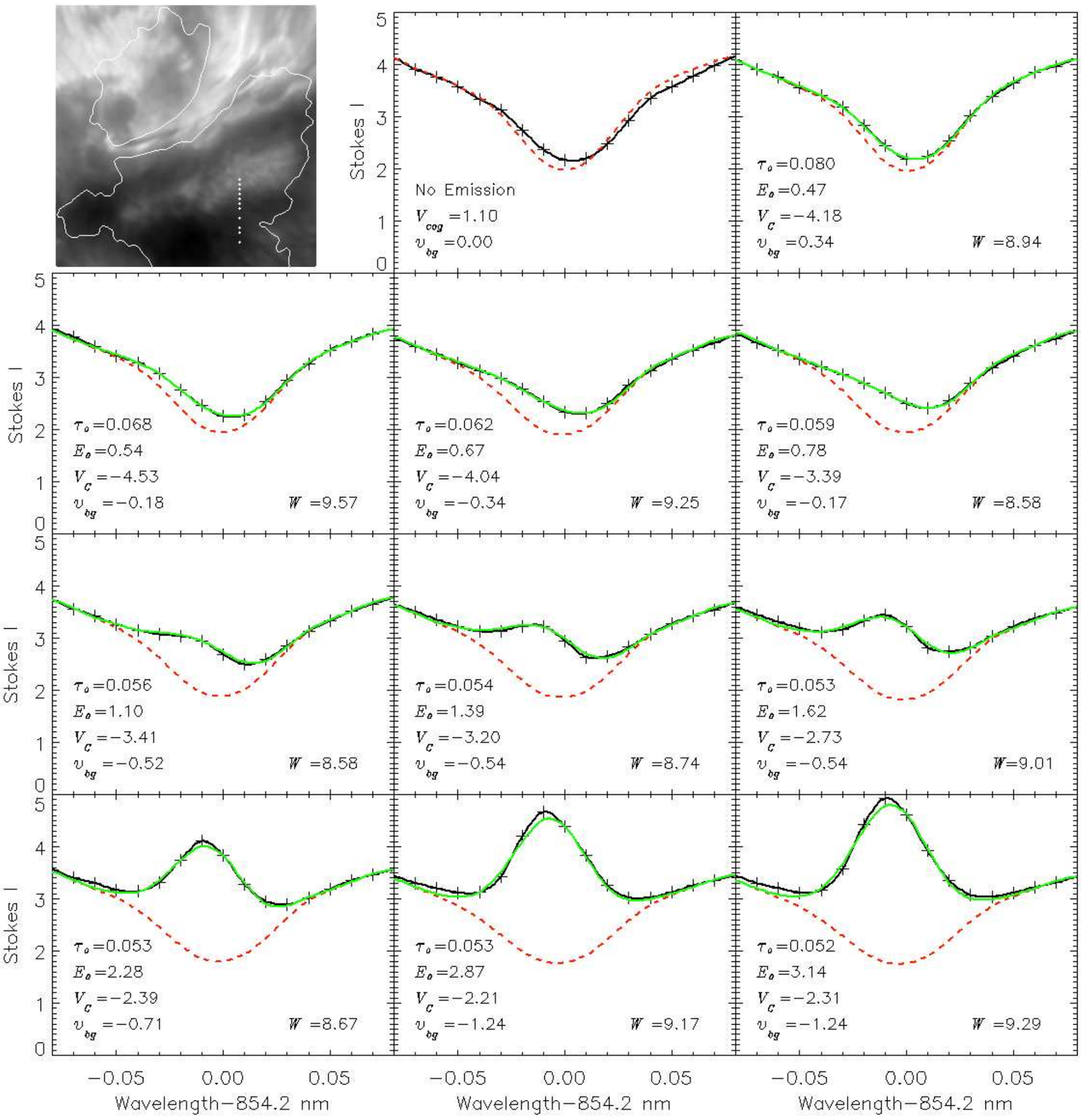}}
 \caption[]{Different spectral profiles from a vertical cut (white small plus symbols
  in top-left panel: line-core filtergram of the third scan) in the quiet umbra when
  an umbral flash is seen. Observed profile: plus symbols and black solid lines;
  background incident profile: red dashed lines; fitted profiles: green solid lines.
  The lowest white plus symbol in top-left panel corresponds to the top-middle panel
  and the highest white plus symbol corresponds to the bottom-right panel.
 }
 \label{bestfit_q}
\end{figure}

\begin{figure} %Fig.~17
 \centerline{\includegraphics[width=\textwidth,clip=]{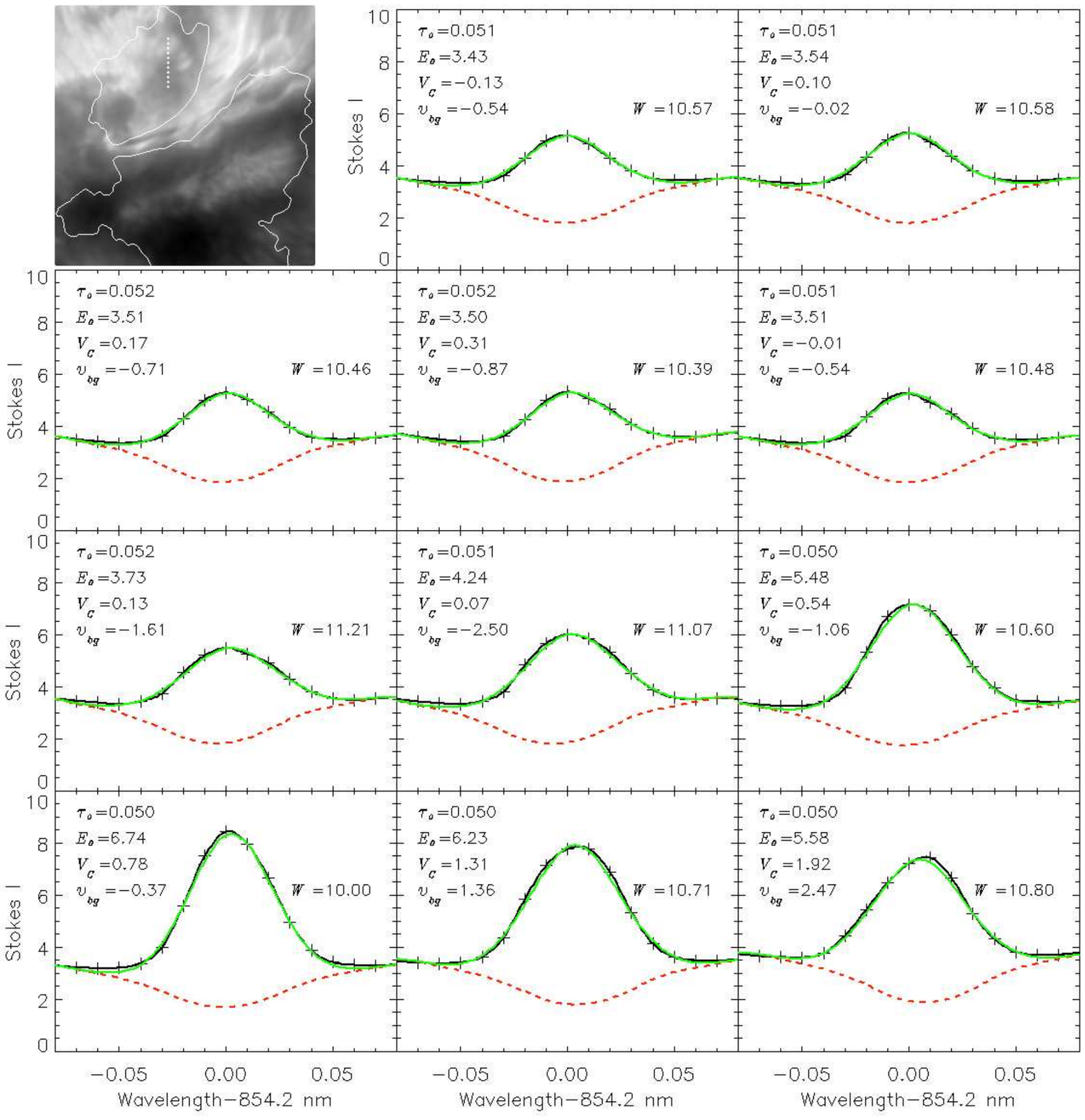}}
 \caption[]{Same as in Fig.~\ref{bestfit_q} but for different spectral profiles from a vertical
 cut (plus symbols in top-left panel) in the active umbra.
 }
 \label{bestfit_a}
\end{figure}

For a
better illustration of the efficiency of thin cloud model to
retrieve atmospheric physical parameters, some pixels with different
spectral profiles from two vertical cuts, one in the quiet umbra
when an umbral flash is seen and the other in the active umbra at
the same time, were selected. The fitted profiles and their
corresponding atmospheric physical parameters are given in
Figs.~\ref{bestfit_q} \& \ref{bestfit_a}, respectively. Top middle
panel in Fig.~\ref{bestfit_q} shows an observed profile with a
typical absorption shape that does not show emission signature. This
profile shows a center-of-gravity velocity $V_{cog}$ of
1.1~km~s$^{-1}$.

%----------------------------------------------------------------------
%----------------------------------------------------------------------
%----------------------------------------------------------------------
%----------------------------------------------------------------------
%----------------------------------------------------------------------
%----------------------------------------------------------------------
%----------------------------------------------------------------------
\begin{figure} %Fig.~18
 \centerline{\includegraphics[angle=90, width=\textwidth,clip=]{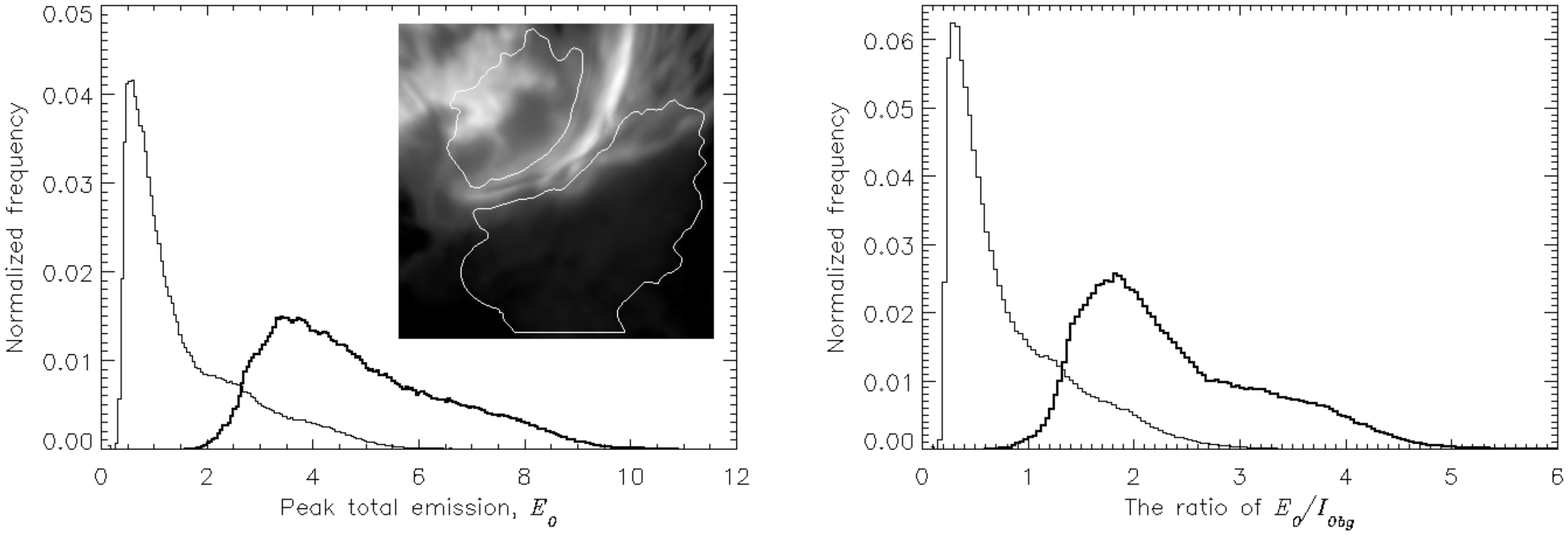}}
 \caption[]{Left panel: histograms of the peak total emission $E_{0}$ in the quiet
 umbra (thin solid line) and in the active umbra (thick solid line). The inset image
 is the time average of $E_{0}$. Pixels in quiet and active umbra used to construct
 the histograms are enclosed by white contours. Right panel: corresponding histograms
 of the ratios of the peak total emission $E_{0}$ to the corresponding line-core
 intensity of the incident background profile $I_{0bg}$ in both active (thick solid
 line) and quiet umbra (thin solid line).
 }
 \label{hist_e0}
\end{figure}
%----------------------------------------------------------------------
\begin{figure} %Fig.~19
 \centerline{\includegraphics[angle=90, width=\textwidth,clip=]{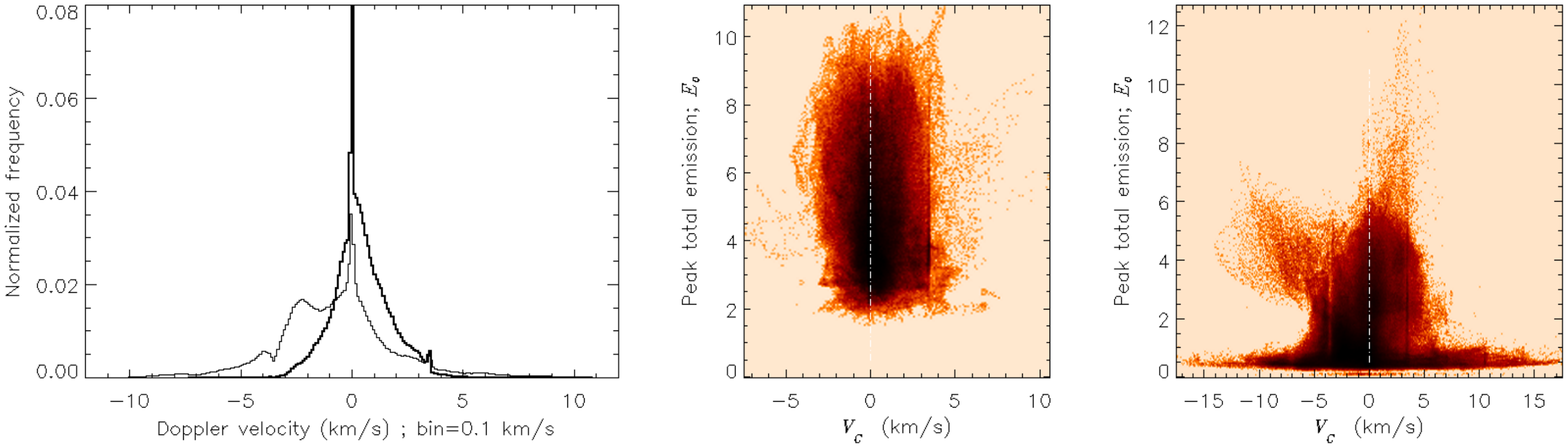}}
 \caption[]{Left panel: histograms of Doppler velocities ($V_{C}$) in the quiet umbra
 (thin solid line) and in the active umbra (thick solid line). Negative Doppler
 velocities show up-flows. Middle panel: scatter plot of $E_{0}$ versus $V_{C}$ for
 the active umbra. Right panel: scatter plot of $E_{0}$ versus $V_{C}$ for quiet umbra.
 The color map scales the population/frequency of each point; darker region means more
 frequency. Pixels in the quiet and active umbra used to construct the histograms and
 scatter plots are the same as those defined in Fig.~\ref{hist_e0}.
 }
 \label{e0_vc}
\end{figure}
%----------------------------------------------------------------------

\section{Conclusions}\label{conclusion}
According to the results of the solution of the radiative transfer
equation based on thin cloud model, we can compare the dynamical
atmospheric parameters of quiet and active umbra; the active umbra
shows a steady emission in Ca~\texttt{II}~854.2~nm line as well as
in H$\alpha$ line, with a high source function and equivalently,
with a large peak total emission, as can be seen in the left panel
of Fig.~\ref{hist_e0}. On the other hand, pixels inside the quiet
umbra show emissions only when an umbral flash is propagating
through the umbra. During the propagation of an umbral flash, the
peak total emission of pixels inside the quiet umbra varies from
smaller values accompanied by higher upflows to larger values with
smaller up- or downflows. The temporal and spatial averages of the
peak total emission in the quiet and active umbra are about 1.54 and
4.82, respectively. The right panel of Fig.~\ref{hist_e0} displays
histograms of the ratio of the peak total emission $E_{0}$ to the
corresponding line-core intensity of the incident background profile
$I_{0bg}$ in both active (thick solid line) and quiet umbra (thin
solid line). These distributions are similar to the corresponding
distribution of $E_{0}$. This ratio gives us a measure showing how
large is $E_{0}$ and how strong is the spectral emission.

Histograms of resulting Doppler velocities are shown in
Fig.~\ref{e0_vc} (left panel). It can be seen from the figure that
while upflows are dominant in the quiet umbra, the active umbra is
connected mostly with downflows.

The scatter plot of $E_{0}$ versus $V_{C}$ for the quiet umbra
(right panel of Fig.~\ref{e0_vc}) shows two different parts: very
large values of $E_{0}$ are related to downflows with Doppler
velocities less than 5~km~s$^{-1}$; while, a considerable population
of pixels with moderate $E_{0}$($<7$) shows up-flows with large
Doppler velocities up to 15 km~s$^{-1}$. A detailed description of
the obtained velocity maps and the possible interpretation of the
velocity distributions will be discussed elsewhere.

\section*{Acknowledgment}
H. Hamedivafa thanks the Imam Khomeini International University for
supporting his sabbatical stay at the Astronomical institute of the
Czech Academy of Sciences. The work was also supported by the Czech
Science Foundation under the grant 14-04338S. The Swedish 1-m Solar
Telescope is operated on the island of La Palma by the Institute for
Solar Physics of Stockholm University in the Spanish Observatorio
del Roque de los Muchachos of the Instituto de Astrof\'{\i}sica de
Canarias. We acknowledge financial support by the Spanish Ministerio
de Econom\'{\i}a y Competitividad through projects
AYA2012-39636-C06-05 and ESP2013- 47349-C6-1-R, including a
percentage from European FEDER funds. Also, H. Hamedivafa thanks P.
Heinzel and J. \v{S}tepan for their helpful discussions and
suggestions.

%Wang \& Zirin \cite{wz92}
%e.g., \v{S}vanda \cite{svanda07} and Verma \& Denker \cite{verma11}
%Roudier et al. \cite{roud99}
%(e.g., \cite{svanda07,sob99b,rieutord01})
%(see Sect.~\ref{zfr}).
%Figs.~\ref{gvecflow} \& \ref{gspeed}, arrows~1~\&~2
%Bellot Rubio \cite{bell03}
%Borrero \& Ichimoto \cite{bor_ich11}
%\newpage
%\mbox{}
%\newpage

\end{document}